\documentclass[prb,twocolumn,showpacs]{revtex4}
\usepackage{graphicx}
\usepackage{amsmath}
\usepackage{amssymb}
\usepackage{dcolumn}
\usepackage{float}
\usepackage{bm}
\usepackage[breaklinks=true,colorlinks,citecolor=blue,linkcolor=blue,urlcolor=blue]{hyperref}

\newcommand{\nn}{\nonumber}

\DeclareMathAlphabet{\bi}{OML}{cmm}{b}{it}

\def\be{\begin{equation}}
\def\ee{\end{equation}}
\def\bearr{\begin{eqnarray}}
\def\eearr{\end{eqnarray}}
\def\la{\langle}
\def\ra{\rangle}

\begin{document}
\title{Magnetotransport properties of the $\alpha$-T$_3$ model}
\bigskip

\author{Tutul Biswas}
\email{tbtutulm53@gmail.com}
\author{Tarun Kanti Ghosh
}
\email{tkghosh@iitk.ac.in}
\normalsize
\affiliation
{$\color{blue}{^\ast}$Department of Physics, Vivekananda Mahavidyalaya, Burdwan-713103, India\\
$\color{blue}{^\dagger}$Department of Physics, Indian Institute of Technology-Kanpur,
Kanpur-208 016, India}
\date{\today}

\begin{abstract}
Using the well-known Kubo formula, we evaluate magnetotransport quantities like
the collisional and Hall conductivities of the $\alpha$-T$_3$ model. The 
collisional conductivity exhibits a series of peaks at strong magnetic 
field. Each of the conductivity peaks for $\alpha=0$ (graphene) splits 
into two in presence of a finite $\alpha$. This splitting occurs due to 
a finite phase difference between the contributions coming from the two 
valleys. The density of states is also calculated to explore the origin
of the splitting of conductivity peaks. As $\alpha$ approaches $1$, the right split 
part of a conductivity 
peak comes closer to the left split part of the next conductivity peak. 
At $\alpha=1$, they merge with each other to produce a new series of the 
conductivity peaks. On the other hand, the Hall conductivity undergoes a 
smooth transition from $\sigma_{yx}=2(2n+1)e^2/h$ to $\sigma_{yx}=4ne^2/h$ with
$n=0,1,2,...$ as we tune $\alpha$ from $0$ to $1$. For intermediate $\alpha$,
we obtain the Hall plateaus at values $0,2,4,6,8,...$ in units of $e^2/h$.  
\end{abstract}



\maketitle


\section{Introduction}
The signatures of the Dirac physics in realistic systems have been established 
after the phenomenal discovery of graphene monolayer \cite{Grph_dis,grph}. 
Graphene, a strictly two-dimensional sheet of carbon atoms arranged on a honeycomb 
lattice (HCL) structure, exhibits low-energy excitations which are linear in momentum. 
The quasi-particles in graphene obey pseudospin $S=1/2$ Dirac-Weyl equation. 
The conduction band meets with the valence band at the six corner points of 
the hexagonal first Brillouin zone (BZ), known as Dirac points. A number of 
fascinating physical phenomena have been emerged in graphene in recent years. 
Unconventional integer quantum Hall effect \cite{QHE1,QHE2,QHE3,QHE4} is one 
of them in which quantization occurs due to the quantum 
anomaly\cite{QHE2} of the zero-energy Landau level.

On the other hand, there exists an analogous lattice, the so-called dice or 
$T_3$-lattice \cite{dice1,dice2} in which quasi-particles are characterized by 
the Dirac-Weyl equation with an enlarged pseudospin $S=1$. An unit cell of the 
$T_3$-lattice consists of three inequivalent lattice sites. Two of these, usually 
known as {\it rim} sites, are situated at the corner points of HCL alternatively. 
Both the  {\it rim} sites are connected to the three nearest neighbors (NNs). 
The rest lattice site is called {\it hub} site. It is located at the center of 
HCL and is connected to six NNs. The low-energy excitations near the Dirac points 
consist of three energy branches in which two are linear in momentum, known as
conic band. The non-dispersive third energy branch is usually termed as flat 
band. All the six band-touching points in the first  BZ lie on the flat band.

The $T_3$-lattice, belongs to bipartite class, has been extensively investigated 
within the context of topological localization \cite{dice1,dice2}, 
magnetic frustration \cite{frust1,frust2}, Rashba spin-orbit interaction induced effects 
\cite{dice_SOI1}, Klein tunneling \cite{dice_Klein}, plasmon\cite{plasm} etc. The existence of 
$T_3$-model has been proposed recently using ultra-cold atoms \cite{dice_opt}. It is 
also possible to build a $T_3$-lattice by growing trilayer structures of cubic 
lattices in the $(111)$ direction \cite{dice_grow}. Recently, there has been a 
growing interest on the lattices which are described by the generalized Dirac-Weyl 
equation with arbitrary pseudospin $S$ \cite{dice_S1,dice_S2,dice_S3}.

In addition to the $T_3$-lattice, there is a modified lattice, 
known as $\alpha$-$T_3$ model \cite{dice_alph}, in which
the hopping strength between the {\it hub} site and one of the {\it rim} 
sites is proportional to the parameter $\alpha$. A continuous tuning of $\alpha$ 
demonstrates the crossover between a HCL ($\alpha=0$) and a $T_3$-lattice ($\alpha=1$). 
With appropriate doping \cite{dice_alph2} a Hg$_{1-x}$Cd$_x$Te quantum well can be 
mapped onto a $\alpha$-$T_3$ model with an effective $\alpha=1/\sqrt{3}$. 
The continuous evolution of $\alpha$ is associated with the Berry phase of the system 
and has an enormous effect on the orbital magnetic response. Particularly, the orbital 
susceptibility\cite{dice_alph} of the system changes from dia- to paramagnetic
behavior as one continuously tunes $\alpha$ from $\alpha=0$ (HCL) to $\alpha=1$ ($T_3$). 
A number of physical observables including
DC Hall conductivity\cite{dice_Berry}, dynamical optical conductivity\cite{dice_Berry},
and magneto-optical conductivity\cite{dice_MagOP1,dice_MagOP2} of a $\alpha$-$T_3$ model have been
studied recently and the associated behaviors have also been linked with the Berry phase.

In this work, we study the transport properties of the $\alpha$-$T_3$ model in a 
transverse magnetic field within linear response regime. We use Kubo formalism 
to understand the behavior of the collisional and Hall conductivities with various 
parameters like electron density and magnetic field. In the strong field regime, the 
collisional conductivity is described by a number of peaks. For a finite $\alpha$, the 
peaks arising in the longitudinal conductivity split because the contributions coming 
from different valleys are different in a  phase. A finite $\alpha$ introduces additional 
plateaus exactly at the midway between the Hall plateaus obtained in the case of 
graphene ($\alpha=0$). We observe a transition in the Hall conductivity from 
$\sigma_{yx}=2(2n+1)e^2/h$ (for $\alpha=0$) to $\sigma_{yx}=4ne^2/h$ (for $\alpha=1$) 
with $n=0,1,2,..\,.$. 

This paper is presented in the following way. In section II, 
we discuss basic informations of the system including Hamiltonian, eigen values, 
wave functions, and velocity. Various magnetotransport related quantities are derived in 
section. III. Section IV includes the analysis of the results obtained.
We summarize main outcomes of this paper in section V.

\section{Preliminary informations of the system}

\subsection{Hamiltonian}

Within the framework of the $\alpha$-T$_3$ model, there exists three 
atoms, namely, P, Q, and R in a unit cell as shown in Fig. 1(a). 
The atoms P and Q form a honeycomb 
lattice structure analogous to graphene with a hopping amplitude $t$. The 
atom R is connected to the atom P via a hopping amplitude $\alpha t$. The parameter 
$\alpha$ is the key element of this model. The magnitude of $\alpha$ varies from
$0$ to $1$. The two limiting values of $\alpha$, namely, $\alpha=0$ and $\alpha=1$ 
represent graphene and dice lattice, respectively.

\begin{figure}[h!]
\begin{minipage}[b]{\linewidth}
\centering
 \includegraphics[width=5.5cm, height=4.5cm]{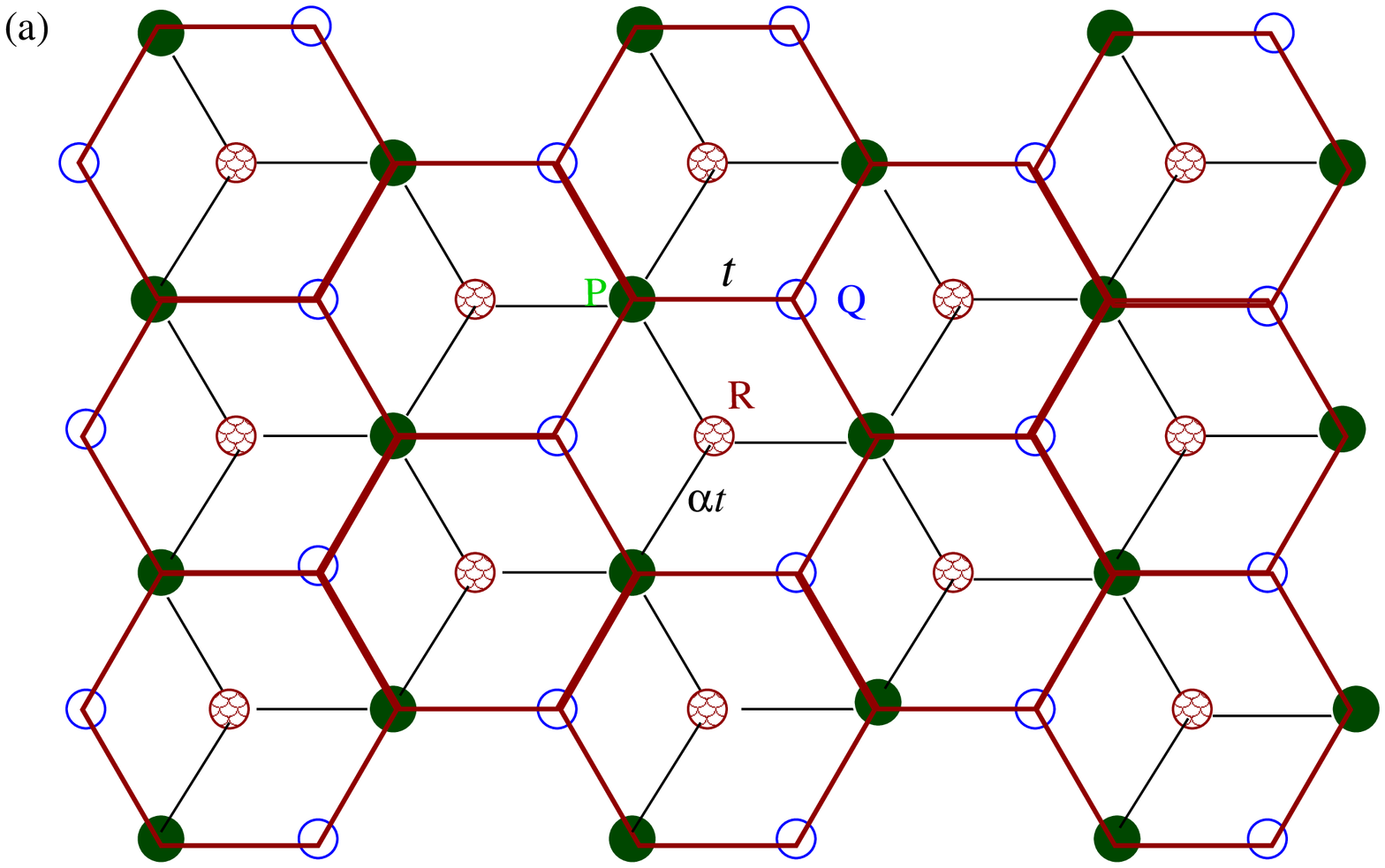}
\end{minipage}
\vspace{1em}

\begin{minipage}[b]{\linewidth}
\centering
\includegraphics[width=5.5cm, height=5cm]{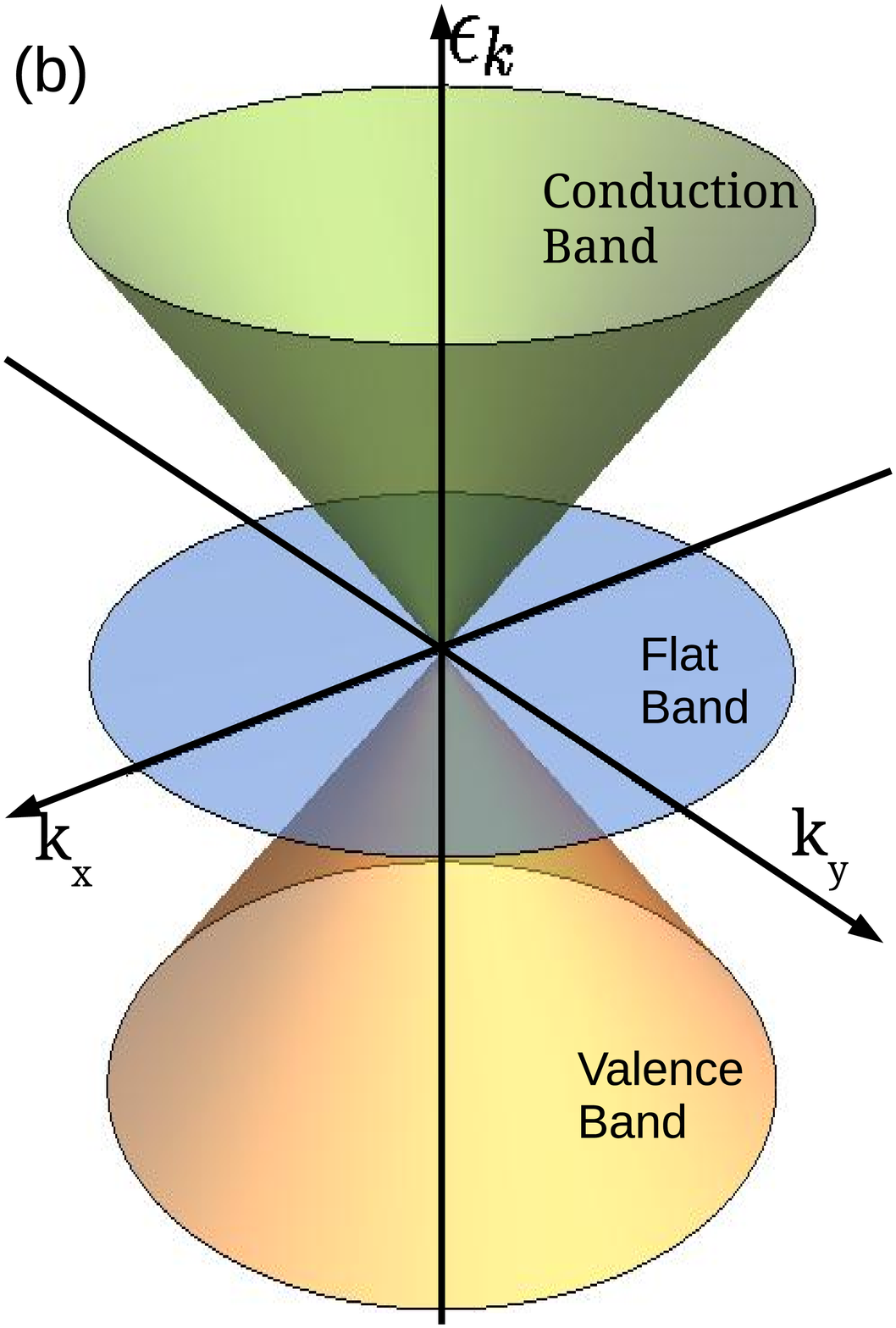}
\end{minipage}

\caption{Sketch of (a) the geometric structure and
 (b) energy spectrum of a $\alpha$-T$_3$ lattice.}
\end{figure}

Within the tight-binding approximation the low-lying energy states near 
the Dirac point in a particular valley  are described by the Hamiltonian\cite{dice_alph}

\begin{eqnarray}
H({\bf p})=
\begin{pmatrix}
    0      &  f_{\bf p}\cos\phi &   0 \\
    f_{\bf p}^\ast\cos\phi  &  0  &  f_{\bf p}\sin\phi \\
    0     &  f_{\bf p}^\ast\sin\phi  &  0
\end{pmatrix},
\end{eqnarray}
where $f_{\bf p}=v_F(\zeta p_x-ip_y)$ with $v_F$ being the Fermi velocity. 
The valley index $\zeta=\pm1$ represents ${\rm K}$ and ${\rm K}^\prime$ valley i. e. two
inequivalent Dirac points in the first BZ, 
respectively. The angle $\phi$ is connected to the parameter $\alpha$ via $\alpha=\tan\phi$.

In presence of an external magnetic field ${\bf B}=B\hat{z}$, transverse to the 
crystal plane, we make the following Pierls substitution ${\bf \Pi}={\bf p}+e{\bf A}$, 
where the vector potential ${\bf A}$ is chosen in the Landau gauge as ${\bf A}=(-By,0,0)$.
Hence, the Hamiltonian near the Dirac point in the ${\rm K}$-valley takes the following form
\begin{eqnarray}\label{HamK}
H_{\bf \rm K}=\gamma_B
\begin{pmatrix}
    0      &  \cos\phi\,\hat{a} &   0 \\
    \cos\phi\,\hat{a}^\dagger  &  0  &  \sin\phi\,\hat{a} \\
    0     &  \sin\phi\,\hat{a}^\dagger  &  0
\end{pmatrix},
\end{eqnarray}
where $\gamma_B=\sqrt{2}\hbar v_F/l_0$ with $l_0=\sqrt{\hbar/(eB)}$ 
being the magnetic length. The annihilation and creation operators are 
given by $\hat{a}=v_F\Pi_-/\gamma_B$ and $\hat{a}^\dagger=v_F \Pi_+/\gamma_B$, 
respectively, with $\Pi_\pm=\Pi_x \pm i\Pi_y$. The operators do obey the commutation 
relation $[\hat{a},\hat{a}^\dagger]=1$ and the actions of
them on the Fock states $\vert n\ra$ are the following: 
$\hat{a}\vert n\rangle=\sqrt{n}\vert n-1\rangle$ and
$\hat{a}^\dagger\vert n\rangle=\sqrt{n+1}\vert n+1\rangle$.
The Hamiltonian corresponding to the ${\bf \rm K}^\prime$-valley is 
obtained through the substitution $\hat{a}\rightarrow -\hat{a}^\dagger$.

\subsection{Conic band}
In the absence of magnetic field the conic band of the $\alpha$-$T_3$ model consists of
conduction and valence bands which disperse linearly with momentum (see Fig. 1(b)). A perpendicular
magnetic field causes to break the continuous energy branches into Landau levels.

On diagonalizing Eq. (\ref{HamK}), the energy spectrum of the system can be 
obtained in the following form
\begin{eqnarray}\label{Energy}
\varepsilon_{n,\zeta}^\lambda=\lambda \, \gamma_B\sqrt{n+\chi_\zeta},
\end{eqnarray}
where
$n=0,1,2,...$ and $\lambda=\pm1$ denotes the conduction band and valence 
band, respectively. The quantity $\chi_\zeta$ depends on the valley index 
$\zeta$ through $\chi_\zeta=[1-\zeta\cos(2\phi)]/2$.

The eigenfunction for $n>0$ corresponding to the K-valley is given by

\begin{eqnarray}
\Psi_{n,k_x}^{\lambda, {\rm K}}({\bf r})=\frac{1}{\sqrt{2}}\left(
\begin{array}{c}
\frac{\sqrt{n(1-\chi_+)}}{\sqrt{n+\chi_+}} \Phi_{n-1}(y)\\
\lambda \Phi_n(y)\\
\frac{\sqrt{(n+1)\chi_+}}{\sqrt{n+\chi_+}}\Phi_{n+1}(y)
\end{array}\right)\frac{e^{ik_xx}}{\sqrt{2\pi}}.
\end{eqnarray}
Here, $\Phi_n(y)=\sqrt{1/(2^nn!\sqrt{\pi}l_0)} e^{-(y-y_0)^2/(2l_0^2)}
H_n[(y-y_0)/l_0]$ with $y_0=l_0^2k_x$ is the usual harmonic oscillator wave function.

For $n=0$ the eigenfunction is given by

\begin{eqnarray}
\Psi_{0,k_x}^{\lambda,{\rm K}}({\bf r})=\frac{1}{\sqrt{2}}\left(
\begin{array}{c}
0\\
\lambda\Phi_0(y)\\
\Phi_1(y)
\end{array}\right)\frac{e^{ik_xx}}{\sqrt{2\pi}}.
\end{eqnarray}

\subsection{Flat band}
In addition to the spectrum (Eq. (\ref{Energy})), there exist a 
non-dispersive energy band
$\varepsilon_{n}^{\rm F}=0$ $\forall \, n$, known as flat-band.

The corresponding eigenfunctions for the K-valley can be obtained as

\begin{eqnarray}
\Psi_{n,k_x}^{{\rm F}, {\rm K}}({\bf r})=\left(
\begin{array}{c}
-\frac{\sqrt{(n+1)\chi_+}}{\sqrt{n+\chi_+}}\Phi_{n-1}(y)\\
0\\
\frac{\sqrt{n(1-\chi_+)}}{\sqrt{n+\chi_+}}\Phi_{n+1}(y)
\end{array}\right)\frac{e^{ik_xx}}{\sqrt{2\pi}}
\end{eqnarray}
and
\begin{eqnarray}
\Psi_{0,k_x}^{{\rm F},{\rm K}}({\bf r})=\left(
\begin{array}{c}
0\\
0\\
\Phi_0(y)
\end{array}\right)\frac{e^{ik_xx}}{\sqrt{2\pi}},
\end{eqnarray}
for $n>0$ and $n=0$, respectively.

The eigenfunctions corresponding to the K$^\prime$-valley are given 
in the Appendix A.

\subsection{Velocity operators}
The components of the velocity for the ${\rm K}$-valley can be obtained 
in the following matrix form
\begin{eqnarray}
v_x=\frac{\partial{H}}{\partial{p_x}}=v_F
\begin{pmatrix}
    0      &  \cos\phi &   0 \\
    \cos\phi  &  0  &  \sin\phi \\
    0     &  \sin\phi  &  0
\end{pmatrix}
\end{eqnarray}
and

\begin{eqnarray}
v_y=\frac{\partial{H}}{\partial{p_y}}=v_F
\begin{pmatrix}
    0      &  -i\cos\phi &   0 \\
    i\cos\phi  &  0  &  -i\sin\phi \\
    0     &  i\sin\phi  &  0
\end{pmatrix}.
\end{eqnarray}
To find the velocity components for the ${\rm K}^\prime$-valley, we need to make
the replacement $v_x\rightarrow -v_x$ and $v_y\rightarrow v_y$.

\section{Magneto-transport coefficients}
In the regime of linear response theory, where the electric field is weak enough,
we will derive here the analytical expressions for the components of the conductivity
tensor. To do this we would employ the well-known Kubo formalism \cite{Kubo}. 
The diagonal components of the conductivity tensor, known as longitudinal conductivity 
consists of diffusive and collisional contributions. In presence of perpendicular magnetic 
field, the diffusive contribution would give vanishing result since the diagonal elements 
of the velocity matrix are zero. Hence, the contribution in the longitudinal
conductivity entirely comes from the collisional or hopping process. The off-diagonal 
component is usually termed as transverse or the Hall conductivity.    

\subsection{Collisional conductivity}
Within the Kubo formalism, the general expression for the collisional
conductivity is given by\cite{Kubo_coll1,Kubo_coll2,Kubo_coll3,Kubo_coll4}
\begin{eqnarray}
\sigma_{yy}=\frac{\beta e^2}{S}
\sum_{\xi, \xi^\prime} f(\varepsilon_\xi)\{1-f(\varepsilon_{\xi^\prime})\}
W_{\xi \xi^\prime} (y_\xi-y_{\xi^\prime})^2,
\end{eqnarray}
where $\xi\equiv(n,k_x,\lambda)$ represents the set of all quantum numbers, 
$S$ is the area of the sample, $\beta=1/(k_BT)$ with $T$ being the temperature of 
the system, $y_\xi=\langle \xi\vert y\vert \xi\rangle$, and
$f(\varepsilon_\xi)=[e^{\beta(\varepsilon_\xi-\mu)}+1]^{-1}$ is the Fermi-Dirac 
distribution function with $\mu$ as the chemical potential. 
In addition, $W_{\xi \xi^\prime}$ denotes the probability by which an electron 
makes a transition from an initial state $\vert \xi\rangle$ to a final state 
$\vert \xi^\prime\rangle$. In the case of elastic scattering by static impurities,
its expression is given by
\begin{eqnarray}\label{Trans_Prob}
W_{\xi,\xi^\prime}=\frac{2\pi n_{\rm im}}{\hbar S}
\sum_{\bf q} \vert U({\bf q})\vert^2 \vert F_{\xi, \xi^\prime}\vert^2
\delta(\varepsilon_\xi-\varepsilon_{\xi^\prime}),
\end{eqnarray}
where $n_{\rm im}$ is the density of impurities and 
$U({\bf q})$ is the Fourier transform of the screened Coulomb 
potential $U({\bf r}) = e^2 e^{-k_sr}/(4\pi\epsilon_0\epsilon_r r)$ with
$\epsilon_0$, $\epsilon$, and $k_s$ as the free space permittivity,
dielectric constant of the medium, and screened wave vector, respectively.
The expression of $U({\bf q})$ is given by 
$U({\bf q})=e^2/(4\pi\epsilon_0\epsilon_r\sqrt{q^2+k_s^2})$.
Finally, $F_{\xi, \xi^\prime}$ denotes the form factor which is defined as 
$F_{\xi, \xi^\prime}=\langle \xi^\prime\vert e^{i{\bf q}\cdot{\bf r}}\vert\xi\rangle$. 
The square of $F_{\xi, \xi^\prime}$ for the valley $\zeta$ can be obtained as
\begin{eqnarray}\label{Form1}
\Big\vert F_{\xi, \xi^\prime}^\zeta\Big\vert^2 & = 
&\frac{1}{4}\frac{n!}{n^\prime!}u^{n-n^\prime}e^{-u}
\Bigg[\frac{n^\prime(1-\chi_\zeta)}{\sqrt{(n+\chi_\zeta)(n^\prime+\chi_\zeta)}} 
L_{n-1}^{n^\prime-n}(u)\nonumber\\
 & + & \frac{(n+1)\chi_\zeta}{\sqrt{(n+\chi_\zeta)(n^\prime+\chi_\zeta)}}
L_{n+1}^{n^\prime-n}(u) \nonumber\\
& + & \lambda\lambda^\prime L_{n}^{n^\prime-n}(u)\Bigg]^2\delta_{k_x^\prime,k_x+q_x},
\end{eqnarray}
where $u=q^2l_0^2/2$.

To derive an analytical expression for the longitudinal conductivity 
we note that $y_\xi=k_x l_0^2$ and $y_{\xi^\prime}=k_x^\prime l_0^2$. 
With the virtue of $\delta_{k_x^\prime,k_x+q_x}$ given in  Eq. (\ref{Form1}), 
we can write $(y_\xi-y_{\xi^\prime})^2=q_x^2l_0^4$. We now restrict ourselves
to consider only the intra-band ($\lambda^\prime=\lambda$) and intra-level 
($n^\prime=n$) scattering because of the presence of the term 
$\delta(\varepsilon_\xi-\varepsilon_{\xi^\prime})$ in Eq. (\ref{Trans_Prob}). 
With this consideration the form factor for a particular valley $\zeta$ simplifies as 
\begin{eqnarray}\label{Form1}
\Big\vert F_{n,\lambda}^\zeta\Big\vert^2&=&\frac{1}{4}e^{-u}
\Bigg[\frac{n(1-\chi_\zeta)}{n+\chi_\zeta}L_{n-1}(u)+L_{n}(u)\nonumber\\
&+&\frac{(n+1)\chi_\zeta}{n+\chi_\zeta}
L_{n+1}(u)\Bigg]^2 \delta_{k_x^\prime,k_x+q_x}.
\end{eqnarray}

For $n=0$, we have
$\big\vert F_{0,\lambda}\big\vert^2=e^{-u}(1-u/2)^2$ for both the valleys.
Note that both $\big\vert F_{n,\lambda}^\zeta\big\vert^2$ and  
$\big\vert F_{0,\lambda}\big\vert^2$
are independent of $\lambda$.

The sharp Landau levels broaden due to the presence of the impurities in the system. 
Assuming Lorentzian broadening, we may write  
$\delta(\varepsilon_\xi-\varepsilon_{\xi^\prime})=
(1/\pi)\Gamma_0/[(\varepsilon_\xi-\varepsilon_{\xi^\prime})^2+\Gamma_0^2]$, 
where $\Gamma_0$ is the broadening  parameter. It may depend on
magnetic field, quality of samples etc. For intra-level and intra-band scattering we may further write
$\delta(\varepsilon_\xi-\varepsilon_{\xi^\prime})\simeq 1/(\pi\Gamma_0)$. Because of 
the presence of the term $e^{-u}$ in the expressions of 
$\big\vert F_{n,\lambda}^\zeta\big\vert^2$, only small values of $q^2$ are favorable. 
Hence, $U({\bf q})$ can be approximated as 
$U({\bf q})\simeq e^2/(4\pi\epsilon_0\epsilon k_s)\equiv U_0$. We also note that
$\sum_{k_x}\rightarrow g_s S/(2\pi l_0^2)$ with $g_s$ being the spin-degeneracy and 
$\sum_{\bf q}\rightarrow S/(2\pi)^2\int q\,dq\,d\theta$, where $\theta$ is the polar 
angle of ${\bf q}$.

Combining all these, one may arrive at the following formula
\begin{eqnarray}\label{coll_cond1}
\sigma_{yy}&=&\frac{g_s e^2\beta n_{\rm im} U_0^2}{\pi h\Gamma_0 l_0^2}
\sum_{\lambda,\zeta=\pm}\sum_{n=1}^\infty f\big(\varepsilon_{n,\zeta}^\lambda\big)
\Big\{1-f\big(\varepsilon_{n,\zeta}^\lambda\big)\Big\}\nonumber\\
&\times&\int_0^\infty u\Big\vert F_{n\lambda}^\zeta(u)\Big\vert^2\, du.
\end{eqnarray}

Using the orthogonality of the Laguerre polynomials i. e.
$\int_0^\infty e^{-x}L_m(x)L_n(x)\,dx=\delta_{nm}$
and the recurrence relation
$(n+1)L_{n+1}(x)=(2n+1-x)L_n(x)-nL_{n-1}(x)$,
one can do the integration in Eq. (\ref{coll_cond1}).
Finally, we have
\begin{eqnarray}\label{Coll_Cond1}
\sigma_{yy}=g_s\frac{e^2}{h}\frac{\beta n_{\rm im} U_0^2}{4\pi\Gamma_0 l_0^2}
\sum_{\lambda,\zeta}\sum_{n=1}^\infty f\big(\varepsilon_{n,\zeta}^\lambda\big)
\Big\{1-f\big(\varepsilon_{n,\zeta}^\lambda\big)\Big\} I_n^\zeta,
\end{eqnarray}
where 
\begin{eqnarray}
I_n^\zeta&=&(2n-1) \big\vert A_n^\zeta\big\vert^2+(2n+1)+(2n+3) 
\big\vert B_n^\zeta\big\vert^2\nonumber\\
&-&2nA_n^\zeta-2(n+1)B_n^\zeta\nonumber
\end{eqnarray}
with
$A_n^\zeta=n(1-\chi_\zeta)/(n+\chi_\zeta)$ and 
$B_n^\zeta=(n+1)\chi_\zeta/(n+\chi_\zeta)$.

Additionally, the zeroth Landau level would contribute the following 
amount to the conductivity
\begin{eqnarray}\label{Coll_Cond2}
\sigma_{yy}^0=g_s\frac{e^2}{h}\frac{\beta n_{\rm im} U_0^2}{2\pi\Gamma_0 l_0^2}
\sum_{\lambda,\zeta} f\big(\varepsilon_{0,\zeta}^\lambda\big)
\Big\{1-f\big(\varepsilon_{0,\zeta}^\lambda\big)\Big\}.
\end{eqnarray}

\subsection{Hall Conductivity}
Another important quantity of the linear response theory is the Hall conductivity. Its
general expression is given by \cite{Kubo_coll2,Kubo_coll3,Kubo_coll4,Kubo_Hall}
\begin{eqnarray}\label{Hall1}
\sigma_{yx}=\frac{i\hbar e^2}{S}\sum_{\xi,\xi^\prime}
\frac{(f_\xi-f_{\xi^\prime})\la \xi\vert v_x\vert \xi^\prime\ra \la \xi^\prime\vert v_y\vert\xi\ra}
{(\varepsilon_\xi-\varepsilon_{\xi^\prime})(\varepsilon_\xi-\varepsilon_{\xi^\prime}+i\Gamma_0)}.
\end{eqnarray}

Here, $f_\xi \equiv f(\varepsilon_\xi)$.
The expressions for matrix elements of velocity operators i.e. 
$\la \xi\vert v_i\vert \xi^\prime\ra$ with $i=x,y$ are given in the Appendix B.
As those contain $\delta_{k_x^{\prime}, k_x}$, the summation in Eq. (\ref{Hall1}) 
can be simplified as
\begin{eqnarray}
\sum_{\xi,\xi^\prime}\longrightarrow g_s 
\frac{S}{2\pi l_0^2}\sum_{n,n^\prime,\lambda,\lambda^\prime}.\nonumber
\end{eqnarray}

Finite broadening of the Landau levels has been considered for the collisional conductivity. 
In the case of the Hall conductivity, only the transition between different Landau levels 
is important. For sharp levels, the results would be more appropriate. Hence, 
we take $\Gamma_0=0$ for simplicity. Now, Eq. (\ref{Hall1}) becomes
\begin{eqnarray}\label{Hall2}
\sigma_{yx}=g_s\frac{i\hbar e^2}{2\pi l_0^2}\sum_{n,n^\prime \lambda,\lambda^\prime}
\frac{(f_n^\lambda-f_{n^\prime}^{\lambda^\prime})Q_{nn^\prime}^{\lambda\lambda^\prime}}
{(\varepsilon_n^\lambda-\varepsilon_{n^\prime}^{\lambda^\prime})^2},
\end{eqnarray}
where $Q_{nn^\prime}^{\lambda\lambda^\prime}=
\la \Psi_{n}^\lambda\vert v_x\vert \Psi_{n^\prime}^{\lambda^\prime}\ra
\la \Psi_{n^\prime}^{\lambda^\prime}\vert v_y\vert \Psi_{n}^\lambda\ra$. Note that,
the valley index $\zeta$ is omitted from Eq. (\ref{Hall2}). We will calculate 
$\sigma_{yx}$ individually for different valleys and restore the index $\zeta$ later.

It is worthy to mention that two different kind of transitions are possible to occur. 
One is transitions between various states within the conic band. The other type 
is the transition from the flat band to the conic band and vice versa. Let us discuss 
both the contributions one by one.

\subsubsection{Transitions within the Conic Band}
Since the wave function corresponding to $n=0$ and $n>0$ Landau levels are 
different, we expand the summation in Eq. (\ref{Hall2}) explicitly as 
\begin{eqnarray}\label{Hall3}
\sigma_{yx}^{\rm C}&=&D\Bigg[
\sum_{n, \lambda, \lambda^\prime}
\frac{f_n^\lambda-f_0^{\lambda^\prime}}{(\varepsilon_n^\lambda-\varepsilon_0^{\lambda^\prime})^2}
Q_{n0}^{\lambda,\lambda^\prime} + \sum_{n^\prime, \lambda, \lambda^\prime}
\frac{f_0^{\lambda}-f_{n^\prime}^{\lambda^\prime}}
{(\varepsilon_0^\lambda-\varepsilon_{n^\prime}^{\lambda^\prime})^2}
Q_{0n^\prime}^{\lambda,\lambda^\prime}\nonumber\\
& + &\sum_{n,n^\prime \lambda,\lambda^\prime}
\frac{(f_n^\lambda-f_{n^\prime}^{\lambda^\prime})Q_{nn^\prime}^{\lambda\lambda^\prime}}
{(\varepsilon_n^\lambda-\varepsilon_{n^\prime}^{\lambda^\prime})^2}\Bigg],
\end{eqnarray}
where the symbol ${\rm C}$ is used to denote the conic band and $D=ig_s\hbar e^2/(2\pi l_0^2)$.

The explicit expressions of $Q_{n0}^{\lambda\lambda^\prime}$,
$Q_{0n^\prime}^{\lambda\lambda^\prime}$, and $Q_{nn^\prime}^{\lambda\lambda^\prime}$
are given by
\begin{eqnarray}
Q_{n0}^{\lambda \lambda^\prime}=i\frac{v_F^2}{4}
\Big(\lambda^\prime\frac{\sqrt{n}(1-\chi_\zeta)}
{\sqrt{n+\chi_\zeta}}+\lambda\sqrt{\chi_\zeta}\Big)^2\delta_{n1},
\end{eqnarray}

\begin{eqnarray}
Q_{0n^\prime}^{\lambda \lambda^\prime}
=-i\frac{v_F^2}{4}\Big(\lambda\frac{\sqrt{n^\prime}(1-\chi_\zeta)}
{\sqrt{n^\prime+\chi_\zeta}}+\lambda^\prime\sqrt{\chi_\zeta}\Big)^2
\delta_{n^\prime1},
\end{eqnarray}
and
\begin{eqnarray}
Q_{nn^\prime}^{\lambda\lambda^\prime}
=i\frac{v_F^2}{4}\Big(\vert M_{nn^\prime}^{\lambda\lambda^\prime}\vert^2\delta_{n^\prime n-1}
-\vert N_{nn^\prime}^{\lambda\lambda^\prime}\vert^2\delta_{n^\prime n+1}\Big).
\end{eqnarray}
Here, the explicit forms of $\vert M_{nn^\prime}^{\lambda\lambda^\prime}\vert^2$
and $\vert N_{nn^\prime}^{\lambda\lambda^\prime}\vert^2$ are given in the Appendix B.

To evaluate $\sigma_{yx}^{\rm C}$, we note that four different types of 
the set $(\lambda,\lambda^\prime)$ are possible to include all the transitions. They are 
$(+,+)$, $(-,-)$, $(+,-)$, and $(-,+)$. Considering all the contributions,  
we, finally, arrive at the following expression for the Hall conductivity
\begin{eqnarray}\label{HallT}
\sigma_{yx}^{\rm C}&=&\frac{g_s}{2}\frac{e^2}{h}\sum_{\zeta}\sum_{n=0}^\infty
(n+1)P_{n,n+1}^\zeta\Big[f_n^{+,\zeta}+f_n^{-,\zeta}\nonumber\\
&-&f_{n+1}^{+,\zeta}-f_{n+1}^{-,\zeta}\Big],
\end{eqnarray}
where
\begin{eqnarray}
P_{n,n+1}^\zeta&=&\Bigg[1+\bigg(\frac{\delta_n^\zeta}{\delta_{n+1}^\zeta}\bigg)^2\Bigg] 
\big(1-\chi_\zeta\big)^2+
\Bigg[1+\bigg(\frac{\delta_{n+1}^\zeta}{\delta_n^\zeta}\bigg)^2\Bigg]\chi_\zeta^2\nonumber\\
& + &4\chi_\zeta\big(1-\chi_\zeta\big),
\end{eqnarray}
with $\delta_n^\zeta=\sqrt{n+\chi_\zeta}$.

\subsubsection{Transition between Conic band and Flat band}
The electronic transitions from the flat band to the conic band and vice-versa
produce a finite contribution to the Hall conductivity. Particularly, it is crucial
for low-density where small number of Landau levels contribute to the summation. The 
resulting contribution can be written as
\begin{eqnarray}\label{HallFC}
&\sigma_{yx}^{\rm CF}&=2D\sum_{\lambda^\prime}\Bigg[
\frac{f_0^{\lambda^\prime}-f_0^{\rm F}}{(\varepsilon_0^{\lambda^\prime}-\varepsilon_0^{\rm F})^2}
Q_{00\lambda^\prime}^{\rm FC} + \sum_{n=1}^\infty
\frac{f_0^{\lambda^\prime}-f_n^{\rm F}}{(\varepsilon_0^{\lambda^\prime}-\varepsilon_n^{\rm F})^2}
Q_{0n\lambda^\prime}^{\rm FC} \nonumber\\
& + & \sum_{n^\prime=1}^\infty
\frac{f_{n^\prime}^{\lambda^\prime}-f_0^{\rm F}}
{(\varepsilon_{n^\prime}^{\lambda^\prime}-\varepsilon_0^{\rm F})^2}
Q_{n^\prime 0\lambda^\prime}^{\rm FC}
+\sum_{n\neq n^\prime}^\infty
\frac{f_{n^\prime}^{\lambda^\prime}-f_{n}^{\rm F}}
{(\varepsilon_{n^\prime}^{\lambda^\prime}-\varepsilon_n^{\rm F})^2}
Q_{n^\prime n\lambda^\prime}^{\rm FC}\Bigg],\nonumber\\
\end{eqnarray}   
where
$Q_{n^\prime n\lambda^\prime}^{\rm CF}=
\la \Psi_{n^\prime}^{\lambda^\prime}\vert v_x\vert 
\Psi_n^{\rm F}\ra\la \Psi_n^{\rm F}\vert v_y\vert \Psi_{n^\prime}^{\lambda^\prime}\ra$.
The factor `2' in Eq. (\ref{HallFC}) arises due to the fact that transition 
is possible from the conic band to the flat band and vice-versa.

By calculating all $Q_{n^\prime n\lambda^\prime}^{\rm CF}$ explicitly for 
the valley $\zeta$, Eq. (\ref{HallFC}) can be reduced further into the 
following simplified form
\begin{eqnarray}\label{HallFC2}
\sigma_{yx\,\zeta}^{\rm CF}&=&\frac{g_s}{2}\frac{e^2}{h} 
\Bigg[\Big(2\zeta\frac{1-\chi_\zeta}{1+\chi_\zeta}-1\Big)
\big(f_0^{+\zeta}+f_0^{-\zeta}-2f_0^{\rm F}\big)\nonumber\\
& + & \chi_\zeta(1-\chi_\zeta)\sum_{n=1}^\infty
\frac{(n+2)G_n^\zeta-nG_{n+1}^\zeta}{(n+\chi_\zeta)(n+1+\chi_\zeta)}\Bigg],
\end{eqnarray}
where $G_n^\zeta=f_n^{+\zeta}+f_n^{-\zeta}-2f_n^{\rm F}$. In deriving Eq. (\ref{HallFC2}), 
we have also used the fact $f_n^{\rm F}=f_{n+1}^{\rm F}$ since 
$\varepsilon_n^{\rm F}=0$ $\forall \, n$. By summing up the contributions corresponding to 
the different valleys we would obtain the net Hall conductivity due
to the flat-conic band transitions and this amount adds up to Eq. (\ref{HallT}) in 
order to obtain the total Hall conductivity.

\section{Results and Discussions}
Here, we discuss various features of the collisional and Hall conductivities 
obtained through the numerical evaluation of Eqs. (\ref{Coll_Cond1}), (\ref{Coll_Cond2}), 
(\ref{HallT}), and (\ref{HallFC2}). To do this we use the following parameters: 
$\epsilon_r=2.5$, $k_s=10^8$ m$^{-1}$, and $n_{\rm im}=1.5\times 10^{13}$ m$^{-2}$. 
We also consider $\Gamma_0=0.07\gamma_B$ that means $\Gamma_0 \propto \sqrt{B}$. 
The existence of this behavior of $\Gamma_0$ has been confirmed in Refs.[\onlinecite{QHE1,broad1}].

\begin{figure}[h!]
\begin{center}\leavevmode
\includegraphics[width=160mm,height=80mm]{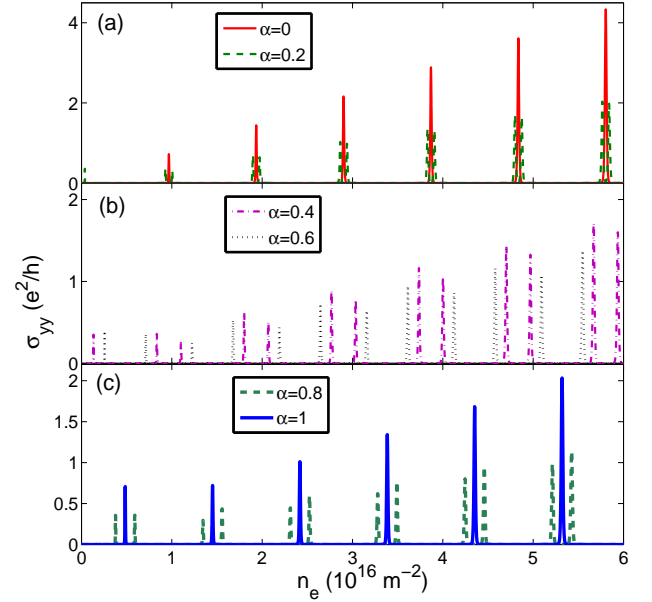}
\caption{(Color online) Plots of the collisional conductivity with the electron density 
for various values of $\alpha$. The magnetic field is fixed to a value $B=10$ T.}
\end{center}
\end{figure}

\begin{figure}[h!]
\begin{center}\leavevmode
\includegraphics[width=175mm,height=75mm]{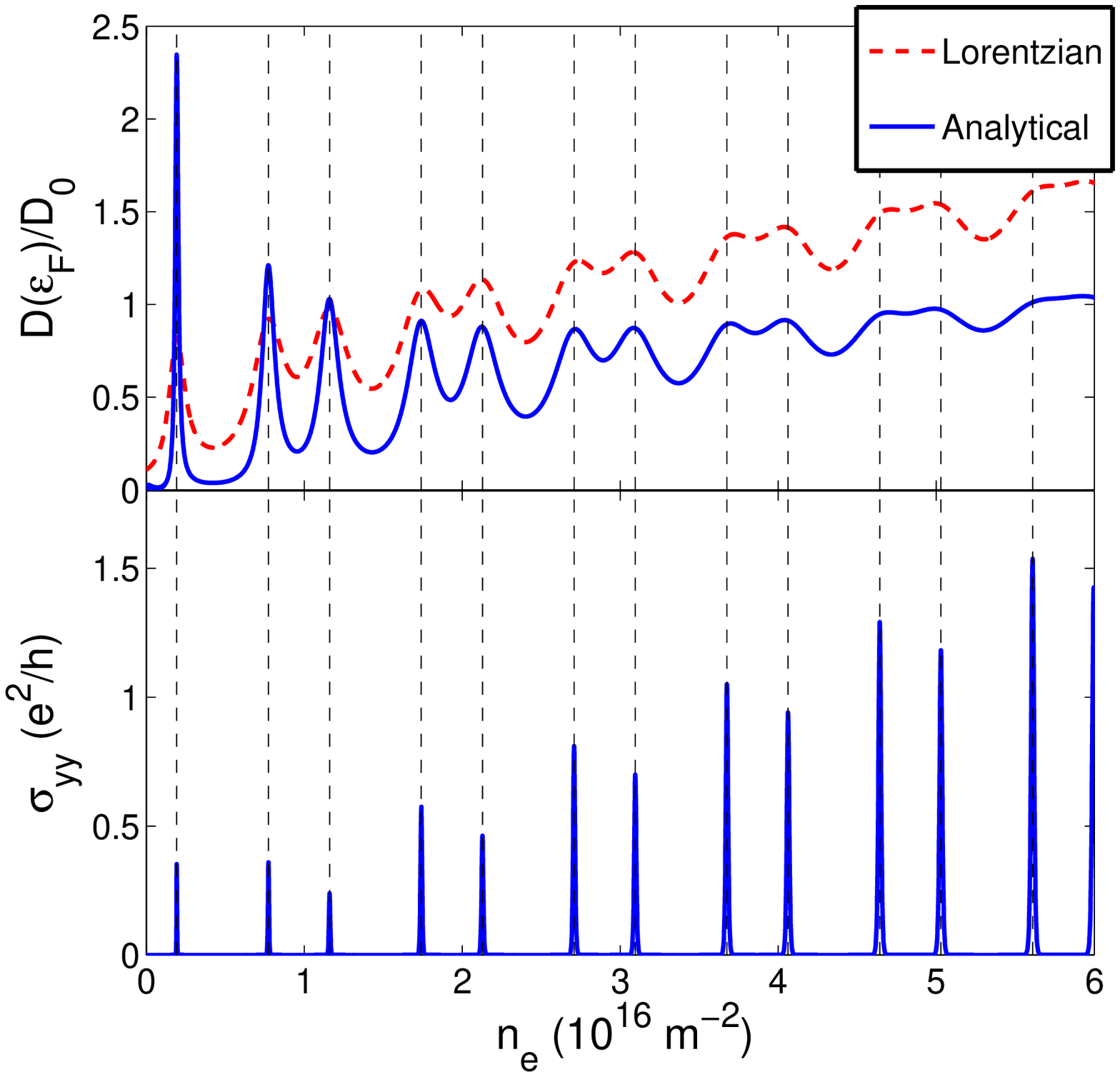}
\caption{(Color online) Plots of DOS and collisional conductivity with the electron 
density for $\alpha=0.5$. The magnetic field is fixed to a value $B=10$ T. 
Here, $D_0=1/(l_0^2\gamma_B)$. In the upper panel, 
the actual value of the solid (blue) line is reduced by a factor $\pi$ for a better visualization.}
\end{center}
\end{figure} 

The variation of the collisional conductivity ($\sigma_{yy}$) with electron 
density ($n_e$) is depicted in Fig. 2 for a constant magnetic field, namely, 
$B=10$ T and different values of $\alpha$, namely, $\alpha=0,0.2,0.4,0.6,0.8,1$. 
When $\alpha=0$ (resembles the case of graphene), $\sigma_{yy}$ displays oscillatory
behavior consisting 
of a number of peaks. This situation changes dramatically as we switch on 
the parameter $\alpha$. For a finite $\alpha$, 
each conductivity peak splits into two peaks which are
unequal in magnitude. The splitting of the peaks can be attributed to the phase difference 
between the contributions arising from different valleys. 
Additionally, a single peak 
appears at very lower density which was absent in the case of graphene. This appearance 
is a direct consequence of the fact that unlike graphene the $n=0$ Landau level is not shared
by the conduction and valence bands. They are distinct in energy separated by a gap of 
$2\sqrt{\chi_\zeta}$ for valley $\zeta$. In other words, the so called ``quantum anomaly''
of $n=0$ level for graphene is absent in  the $\alpha$-$T_3$ model. With the increase of $\alpha$, 
the gap between two split peaks increases. More specifically,
two peaks move in opposite directions. Furthermore, the position of the single peak 
moves towards higher density. Eventually, it merges with the left split-peak of the 
2$^{\rm nd}$ peak when $\alpha$ reaches $1$. Similarly, as $\alpha$ approaches $1$ the 
right split-peak of the 2$^{\rm nd}$ peak merges with the left split-peak of the 
3$^{\rm rd}$ peak and so on. As a result, for $\alpha=1$ we obtain a new set of 
conductivity peaks whose positions are completely different than graphene.
 
The origin of splitting of the conductivity peaks will become
more transparent if we explore the behavior of the density of states (DOS) at 
the Fermi energy.
The motivation behind this is the fact that any transport related quantity is proportional
to the DOS at the Fermi energy.
Generally, the DOS of the Landau levels is given by 
$D(\varepsilon)=\sum \delta(\varepsilon-\varepsilon_{n,\zeta}^\lambda)$. Since the levels are broadened
by impurities, we can replace the $\delta$-function by a Lorentzian distribution as discussed in
section III(A). Hence, the DOS has to be calculated numerically. 
However, it is possible to obtain the following
approximate analytical expression of the DOS (the derivation is given in 
Appendix C),
\begin{eqnarray}\label{DOS_An}
D(\varepsilon_F)=\frac{2\varepsilon_F}{\pi l_0^2\gamma_B^2}\sum_{\zeta}
\Bigg\{1&+&2\sum_{k=1}^\infty 
\exp\Big[-2k\Big(\frac{2\pi\Gamma_0\varepsilon_F}{\gamma_B^2}\Big)^2\Big]\nonumber\\
&\times& \cos\Big[2k\pi\Big(\frac{\varepsilon_F^2}{\gamma_B^2}-\chi_\zeta \Big)\Big]\Bigg\}.
\end{eqnarray}
The DOS at the Fermi energy, calculated numerically and from Eq. (\ref{DOS_An}), 
are depicted in Fig. 3 for a given $\alpha=0.5$. From Fig. 3, we may conclude 
that $D(\varepsilon_F)$ displays similar features as $\sigma_{yy}$
i .e. splitting of peaks and the position of peaks in $D(\varepsilon_F)$ and 
$\sigma_{yy}$ are the same. By considering the most dominant first harmonics ($k=1$ term) 
only in Eq. (\ref{DOS_An}), we may write
$D_\zeta(\varepsilon_F)\sim \cos(\pi^2l_0^2 n_e-2\pi\chi_\zeta)$. 
This clearly indicates that two valleys contribute different amounts to the DOS 
which differ by a phase  $2\pi(\chi_--\chi_+)=2\pi(1-\alpha^2)/(1+\alpha^2)$. 
This phase difference lies entirely behind the splitting of the
conductivity peaks.

\begin{figure}[h!]
\begin{center}\leavevmode
\includegraphics[width=175mm,height=80mm]{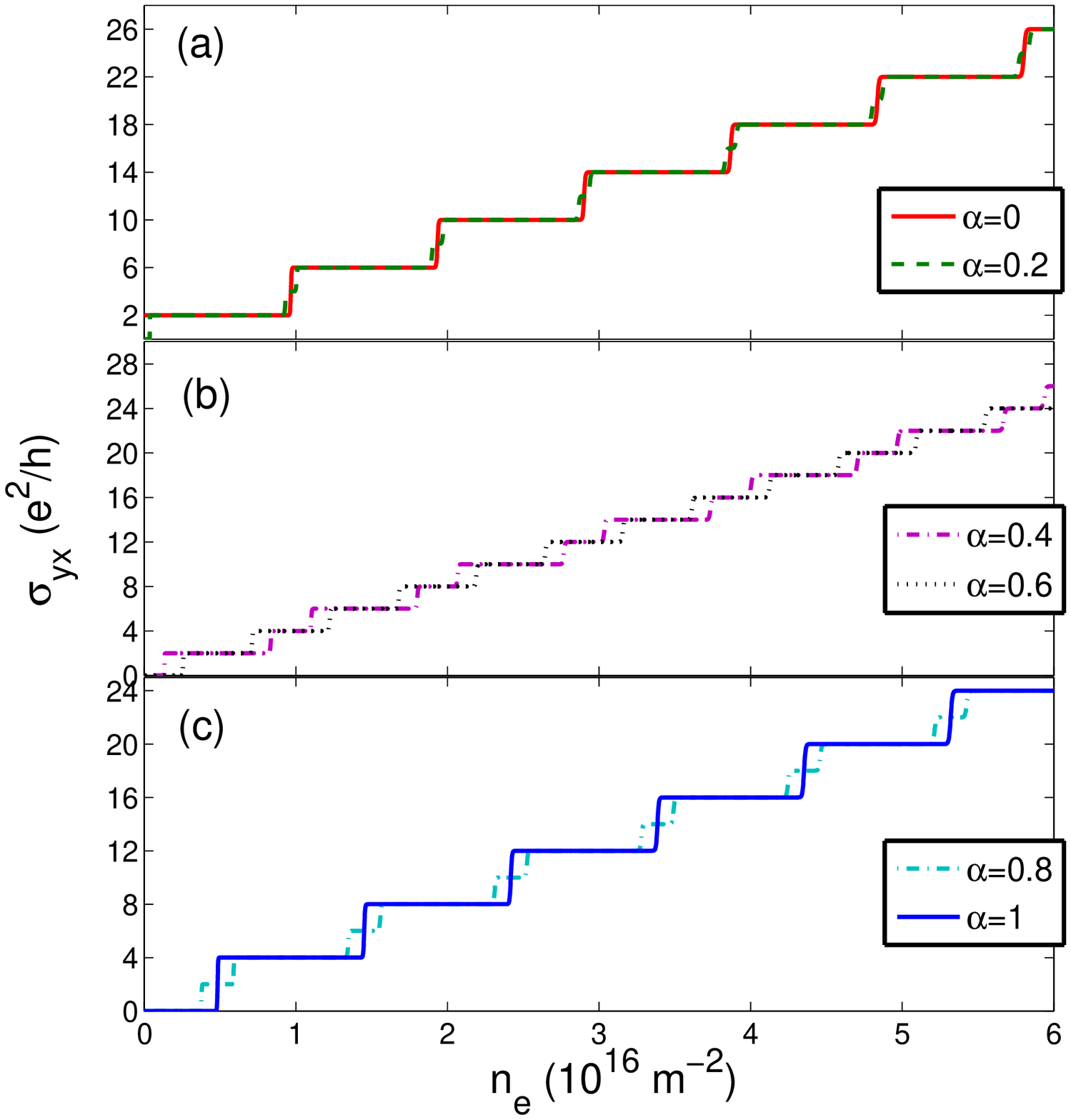}
\caption{(Color online) Plots of the Hall conductivity versus electron density for 
different values of $\alpha$ at fixed $B=10$ T.}
\end{center}
\end{figure}
In Fig. 4, we have shown the behavior of the Hall conductivity as a function 
of the electron density. We fix $B=10$ T and tune $\alpha$ from $0$ to $1$. 
For $\alpha=0$ (graphene), $\sigma_{yx}$ contains a series of Hall plateaus of 
values $2,6,10,...$ in units of $e^2/h$. This type of quantization occurs due 
to the ``quantum anomaly'' of the lowest Landau level\cite{QHE2}. A finite $\alpha$ introduces 
a new series of plateaus situated at the midway between every two plateaus. 
Additionally, a plateau at which $\sigma_{yx}=0$ appears due to the fact the 
lowest Landau-level has non-zero energy and its degeneracy is lifted by a factor `2' 
in presence of a finite $\alpha$. Thus at finite $\alpha$, one obtains the following Hall 
quantization $\sigma_{yx}=2ne^2/h$ with $n=0,1,2,...$\,. The width of each new plateau 
increases and that of each old plateau shrinks with the evolution
of $\alpha$ from $0$ to $1$. Finally, a new series of plateaus of values $0,4,8,...$ in 
units of $e^2/h$ is obtained for $\alpha=1$. Our results are similar to the findings 
of Ref.[\onlinecite{dice_Berry}] in which DC Hall conductivity was indirectly derived from 
magnetization using the Streda formula \cite{streda}. Here, we obtain Hall quantization 
directly via the implementation of the Kubo formalism. In Ref.[\onlinecite{dice_Berry}], the 
contribution of 
flat band was completely ignored due to its zero energy. However, in our treatment the 
transitions between the flat and conic band plays a crucial role, particularly at lower 
density in order to get accurate quantization
of the Hall conductivity.

\begin{figure}[h!]
\begin{center}\leavevmode
\includegraphics[width=160mm,height=60mm]{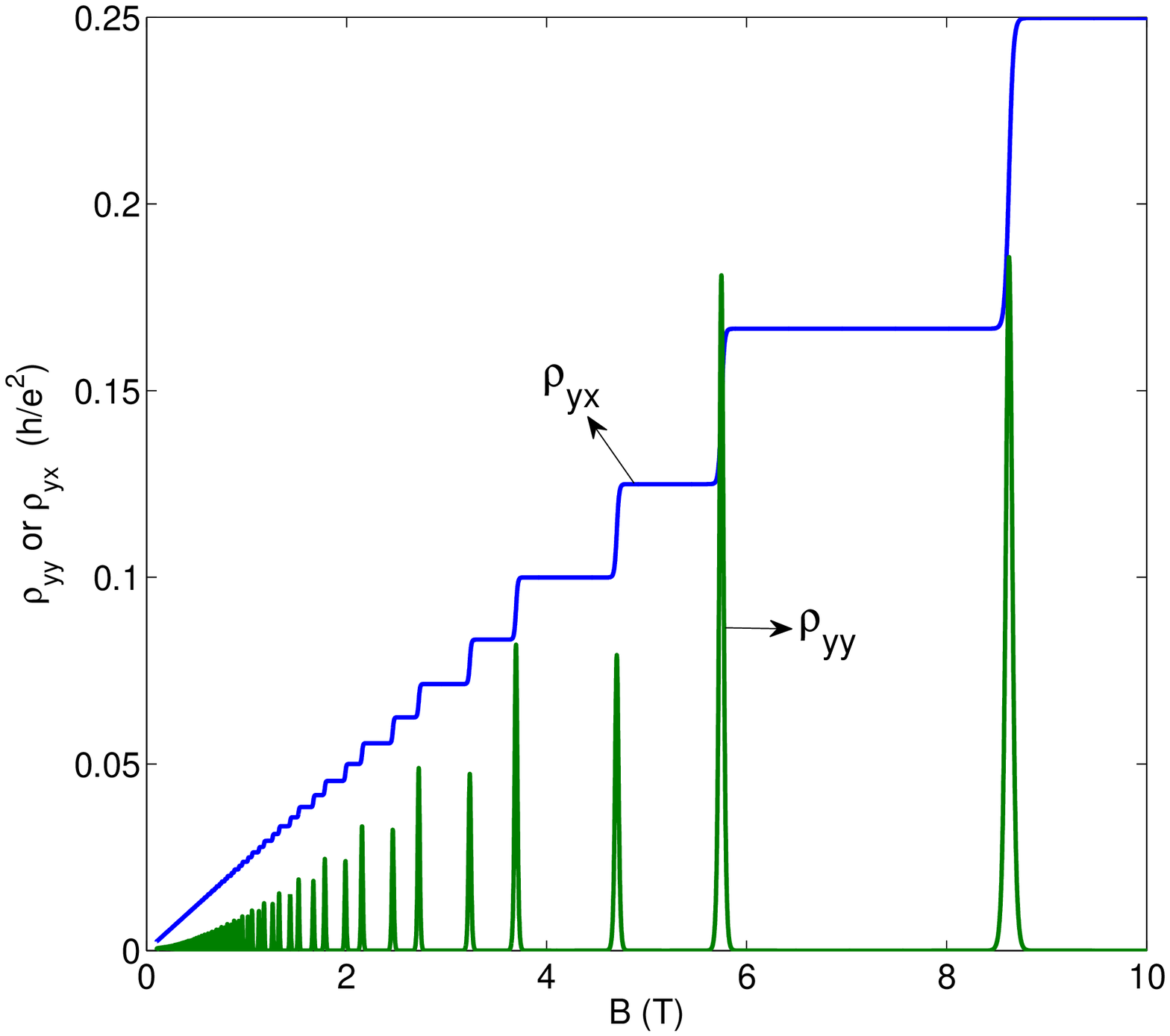}
\caption{(Color online) Plots of $\rho_{yy}$ and $\rho_{yx}$ with magnetic field for $\alpha=0.5$. 
For better visualization, the actual data for $\rho_{yy}$ are enhanced by a factor $20$.}
\end{center}
\end{figure}

The behavior of the resistivity with magnetic field for $\alpha=0.5$ is shown in Fig. 5. 
For every jump in the Hall resistivity from a plateau to the next one,
a peak appears in the longitudinal resistivity.

\section{Summary}
In this work, we have explored the magneto-transport properties of the $\alpha$-T$_3$ 
model by evaluating collisional and the Hall conductivities using the standard Kubo formula. 
At strong magnetic field a number of peaks is appearing in the collisional conductivity. 
The conductivity peaks split into two as a consequence of lifting of the valley degeneracy 
in presence of finite $\alpha$ except at $\alpha=1$. The origin of this splitting has also
been explained through the behavior of density of states, calculated numerically and analytically.
A new series of conductivity peaks 
is obtained for $\alpha=1$ which is different than $\alpha=0$ case.
Like graphene, the Hall conductivity behaves like $\sigma_{yx}=2(2n+1)e^2/h$ with $n=0,1,2,...$ 
for $\alpha=0$. At finite $\alpha$, additional plateaus appear exactly at the mid way 
between every two Hall plateaus.
The width of each new (old) plateau increases (decreases) as $\alpha$
increases from $0$ to $1$. Thus, one obtains the following Hall quantization
$\sigma_{yx}=4ne^2/h$ with $n=0,1,2,...$\, for $\alpha=1$.

\section*{Acknowledgement}
We would like to thank SK Firoz Islam and Alestin Mawrie for useful discussions.

\appendix{}
\section{}
\subsection{Wave functions in K$^\prime$-valley}

For the conic band, the eigen functions corresponding to $n>0$ and $n=0$ are 
given by
\begin{eqnarray}
\Psi_{n,k_x}^{\lambda, {\rm K}^\prime}({\bf r})=\frac{1}{\sqrt{2}}\left(
\begin{array}{c}
\frac{\sqrt{(n+1)\chi_-}}{\sqrt{n+\chi_-}} \Phi_{n+1}(y)\\
-\lambda \Phi_n(y)\\
\frac{\sqrt{n(1-\chi_-)}}{\sqrt{n+\chi_-}}\Phi_{n-1}(y)
\end{array}\right)\frac{e^{ik_xx}}{\sqrt{2\pi}}.
\end{eqnarray}
and 
\begin{eqnarray}
\Psi_{0,k_x}^{\lambda,{\rm K}^\prime}({\bf r})=\frac{1}{\sqrt{2}}\left(
\begin{array}{c}
\Phi_1(y)\\
-\lambda\Phi_0(y)\\
0
\end{array}\right)\frac{e^{ik_xx}}{\sqrt{2\pi}}
\end{eqnarray}

The flat-band wave functions are given by
\begin{eqnarray}
\Psi_{n,k_x}^{\rm F, {\rm K^\prime}}({\bf r})=\left(
\begin{array}{c}
-\frac{\sqrt{n(1-\chi_-)}}{\sqrt{n+\chi_-}}\Phi_{n+1}(y)\\
0\\
\frac{\sqrt{(n+1)\chi_-}}{\sqrt{n+\chi_-}}\Phi_{n-1}(y)
\end{array}\right)\frac{e^{ik_xx}}{\sqrt{2\pi}},
\end{eqnarray}
for $n>0$ and 
\begin{eqnarray}
\Psi_{0,k_x}^{\rm F,{\rm K^\prime}}({\bf r})=\left(
\begin{array}{c}
\Phi_0(y)\\
0\\
0
\end{array}\right)\frac{e^{ik_xx}}{\sqrt{2\pi}},
\end{eqnarray}
for $n=0$.

\section{}
\subsection{Matrix elements of the velocity operators}

The matrix elements of the velocity components for a given valley $\zeta$ can be 
obtained as

\begin{eqnarray}
\big\la \Psi_{nk_x}^{\lambda \zeta}\big\vert v_x\big\vert 
\Psi_{n^\prime k_x^\prime}^{\lambda^\prime \zeta}\big\ra
=\frac{v_F}{2}b\Big(M_{n n^\prime}^{\lambda \lambda^\prime} 
\delta_{n^\prime,n-1}+
N_{n n^\prime}^{\lambda \lambda^\prime}\delta_{n^\prime,n+1}\Big)
\end{eqnarray}

and
\begin{eqnarray}
\big\la \Psi_{n^\prime k_x^\prime}^{\lambda^\prime \zeta}\big\vert v_y\big\vert 
\Psi_{n k_x}^{\lambda\zeta}\big\ra
=\frac{iv_F}{2}b\Big(M_{n n^\prime}^{\lambda \lambda^\prime} 
\delta_{n^\prime,n-1}-
N_{n n^\prime}^{\lambda \lambda^\prime}\delta_{n^\prime,n+1}\Big), 
\end{eqnarray}
where 
\begin{eqnarray}
 M_{n n^\prime}^{\lambda \lambda^\prime}
=\lambda\frac{\sqrt{n^\prime+1}\chi_\zeta}{\sqrt{n^\prime+\chi_\zeta}}
+\lambda^\prime\frac{\sqrt{n}(1-\chi_\zeta)}{\sqrt{n+\chi_\zeta}},
\end{eqnarray}

\begin{eqnarray}
 N_{n n^\prime}^{\lambda \lambda^\prime}
=\lambda\frac{\sqrt{n^\prime}(1-\chi_\zeta)}{\sqrt{n^\prime+\chi_\zeta}}
+\lambda^\prime\frac{\sqrt{n+1}\chi_\zeta}{\sqrt{n+\chi_\zeta}}
\end{eqnarray}
and $b=\delta_{k_x,k_x^\prime}$.

For $0\rightarrow n$ scattering, we have the following matrix elements of $v_x$ and $v_y$
 \begin{eqnarray}
\big\la \Psi_{0k_x^\prime}^{\zeta}\big\vert v_x\big\vert \Psi_{n k_x}^{\lambda\zeta}\big\ra
=\frac{v_F}{2}b\Bigg[\lambda^\prime\frac{\sqrt{n}(1-\chi_\zeta)} 
{\sqrt{n^\prime+\chi_\zeta}} + \lambda\sqrt{\chi_\zeta}\Bigg]\delta_{n,1} \nn
\end{eqnarray}
and 
\begin{eqnarray}
\big\la \Psi_{nk_x}^{\zeta}\big\vert v_y\big\vert \Psi_{0 k_x^\prime}^{\zeta}\big\ra
=\frac{-iv_F}{2}b\Bigg[\lambda^\prime\frac{\sqrt{n}(1-\chi_\zeta)}
{\sqrt{n^\prime+\chi_\zeta}} + \lambda\sqrt{\chi_\zeta}\Bigg]\delta_{n,1}. \nn
\end{eqnarray}

\section{}
\subsection{Calculation of density of states}
Here, we provide an explicit calculation of the density of states (DOS) of the
Landau levels which are broadened by impurities. To calculate DOS, we may start 
from the following expression of the associated self-energy\cite{self1,self2},
\begin{eqnarray}\label{DOS1}
\Sigma^{-}(\varepsilon)=\Gamma_0^2\sum_{n,\zeta} 
\frac{1}{\varepsilon-\varepsilon_{n,\zeta}^\lambda-\Sigma^{-}(\varepsilon)}.
\end{eqnarray}

The imaginary part of $\Sigma^-(\varepsilon)$ is directly related to DOS via
\begin{eqnarray}
D(\varepsilon)={\rm Im}\Bigg[\frac{\Sigma^{-}(\varepsilon)}{\pi^2 l_0^2\Gamma_0^2}\Bigg].
\end{eqnarray}

The summation over $n$ in Eq. (\ref{DOS1}) can be evaluated with the help of
residue theorem i. e. 
$
\sum_n g(n)=-\big\{Sum ~of ~residues~ of~
\pi \cot(\pi z)g(z)~
at ~all~ poles~ of~ g(z)\big\}.$
Inserting $\varepsilon_{n,\zeta}^\lambda$ in Eq. (\ref{DOS1}) we can identify $g(n)$ as
$g(n)=a/(b-c\sqrt{n+\chi_\zeta})$, where $a=\Gamma_0^2$, $b=\varepsilon-\Sigma^{-}(\varepsilon)$,
and $c=\lambda \gamma_B$. Now, the function $g(z)$ has a pole at $z_0=b^2/c^2-\chi_\zeta$.
The residue of $\pi\cot(\pi z)g(z)$ is $-(2ab/c^2)\pi\cot[\pi(b^2/c^2-\chi_\zeta)]$.
Considering the terms which are only linear in $\Sigma^{-}(\varepsilon)$, the self-energy can 
be approximated to the following form
\begin{eqnarray}\label{DOS2}
\Sigma^-(\varepsilon)\simeq \frac{2\pi\Gamma_0^2\varepsilon}{\gamma_B^2}
\cot\Bigg[\pi\Big\{\frac{(\varepsilon^2-2\varepsilon\Sigma^{-}(\varepsilon))}{\gamma_B^2}\Big\}-\chi_\zeta\Bigg]. 
\end{eqnarray}

Separating $\Sigma^-(\varepsilon)$ into real and imaginary parts i. e. 
$\Sigma^-(\varepsilon)=\Delta(\varepsilon)+i\Gamma(\varepsilon)/2$, Eq. (\ref{DOS2})
can be rewritten as
$\Sigma^-(\varepsilon)\simeq (2\pi\Gamma_0^2\varepsilon/ \gamma_B^2)\cot(u-iv)$,
where $u=\pi\Big[\big(\varepsilon^2-2\varepsilon \Delta(\varepsilon)\big)/\gamma_B^2-\chi_\zeta\Big]$
and $v=\pi\varepsilon\Gamma(\varepsilon)/\gamma_B^2$. Now, it is straightforward to obtain
the imaginary part of the self-energy as 
\begin{eqnarray}\label{DOS5}
\frac{\Gamma(\varepsilon)}{2}&=&\frac{2\pi\Gamma_0^2\varepsilon}{\gamma_B^2}
\frac{\sinh(2v)}{\cosh(2v)-\cos(2u)}\nonumber\\
&=& \frac{2\pi\Gamma_0^2\varepsilon}{\gamma_B^2}
\Big[1+2\sum_{k=1}^\infty e^{-2kv}\cos(2ku)\Big].
\end{eqnarray}

In the limit $\pi\varepsilon\Gamma(\varepsilon)/\gamma_B^2<<1$,
 $\Gamma(\varepsilon)$ can be obtained iteratively from Eq. (\ref{DOS5}). After first
iteration, we get $\Gamma(\varepsilon)=4\pi\Gamma_0^2\varepsilon/\gamma_B^2$. Consequently,
the DOS is obtained in the following form
\begin{widetext}
 \begin{eqnarray}
D(\varepsilon)=\frac{2\varepsilon}{\pi l_0^2\gamma_B^2}\sum_{\zeta}
\Bigg\{1+2\sum_{k=1}^\infty \exp\Big[-2k\Big(\frac{2\pi\Gamma_0\varepsilon}{\gamma_B^2}\Big)^2\Big]
\cos\Big[2k\pi\Big(\frac{\varepsilon^2}{\gamma_B^2}-\chi_\zeta \Big)\Big]\Bigg\}.
\end{eqnarray}
\end{widetext}


\begin{thebibliography}{55}

\bibitem{Grph_dis}
K. S. Novoselov, A. K. Geim, S. V. Morozov, D. Jiang, Y. Zhang, S. V. Dubonos,
I. V. Grigorieva, and A. A. Firsov, 
Science {\bf 306}, 666 (2004).

\bibitem{grph}
K. S. Novoselov, A. K. Geim, S. V. Morozov, D. Jiang, M. I.
Katsnelson, I. V. Grigorieva, S. V. Dubonos, A. A. Firsov,
Nature {\bf 438}, 197 (2005).

\bibitem{QHE1}
Y. Zheng and T. Ando, 
Phys. Rev. B {\bf 65}, 245420 (2002).

\bibitem{QHE2}
V. P. Gusynin and S. G. Sharapov, 
Phys. Rev. Lett. {\bf 95}, 146801 (2005).

\bibitem{QHE3}
Y. Zhang, Y.-W. Tan, H. L. Stormer, and P. Kim, 
Nature (London) {\bf 438}, 201 (2005).

\bibitem{QHE4}
K. S. Novoselov, Z. Jiang, Y. Zhang, S. V. Morozov, H. L. Stormer, 
U. Zeitler, J. C. Maan, G. S. Boebinger, P. Kim, and A. K. Geim, 
Science {315}, 1379 (2007).

\bibitem{dice1}
B. Sutherland, 
Phys. Rev. B {\bf 34}, 5208 (1986).

\bibitem{dice2}
J. Vidal, R. Mosseri, and B. Doucot, 
Phys. Rev. Lett. {\bf 81}, 5888 (1998).

\bibitem{frust1}
S. E. Korshunov, 
Phys. Rev. B {\bf 63}, 134503 (2001͒). 

\bibitem{frust2}
M. Rizzi, V. Cataudella, and R. Fazio, 
Phys. Rev. B {\bf 73}, 144511 (2006͒). 

\bibitem{dice_SOI1}
D. Bercioux, M. Governale, V. Cataudella, and V. M. Ramaglia, 
Phys. Rev. Lett. {\bf 93}, 056802 (2004͒); 
Phys. Rev. B {\bf 72}, 075305 (͑2005)͒. 

\bibitem{dice_Klein}
D. F. Urban, D. Bercioux, M. Wimmer, and W. H\"{a}usler, 
Phys. Rev. B {\bf 84}, 115136 (2011).

\bibitem{plasm}
J. D. Malcolm and E. J. Nicol,
Phys. Rev. B {\bf 93}, 165433 (2016).

\bibitem{dice_opt}
D. Bercioux, D. F. Urban, H. Grabert, and W. H\"{a}usler, 
Phys. Rev. A {\bf 80}, 063603 (2009).

\bibitem{dice_grow}
F. Wang and Y. Ran, 
Phys. Rev. B {\bf 84}, 241103 (2011).

\bibitem{dice_S1}
B. Dora, J. Kailasvuori, and R. Moessner, 
Phys. Rev. B {\bf 84}, 195422 (2011).

\bibitem{dice_S2}
Z. Lan, N. Goldman, A. Bermudez, W. Lu, and P. \"{O}hberg, 
Phys. Rev. B {\bf 84}, 165115 (2011).

\bibitem{dice_S3} 
J. D. Malcolm and E. J. Nicol, 
Phys. Rev. B {\bf 90}, 035405 (2014).

\bibitem{dice_alph}
A. Raoux, M. Morigi, J.-N. Fuchs, F. Piechon, and G. Montambaux, 
Phys. Rev. Lett. {\bf 112}, 026402 (2014).

\bibitem{dice_alph2}
J. D. Malcolm and E. J. Nicol, 
Phys. Rev. B {\bf 92}, 035118 (2015).

\bibitem{dice_Berry}
E. Illes, J. P. Carbotte, and E. J. Nicol, 
Phys. Rev. B {\bf 92}, 245410 (2015).

\bibitem{dice_MagOP1}
A. D. Kovacs, G. David,  B. Dora, and J. Cserti,
arXiv:1605.09588. 

\bibitem{dice_MagOP2}
E. Illes, and E. J. Nicol, 
arXiv:1606.00823.


\bibitem{Kubo}
G. M. Eliashberg, 
Sov. Phys.-JETP {\bf 14}, 886 (1962).

\bibitem{Kubo_coll1}
M. Charbonneau, K. M. Van Vliet, and P. Vasilopoulos, 
J. Math. Phys. {\bf 23}, 318 (1982).

\bibitem{Kubo_coll2}
P. Vasilopoulos, 
Phys. Rev. B {\bf 32}, 771 (1985).

\bibitem{Kubo_coll3}
F. M. Peeters and P. Vasilopoulos, 
Phys. Rev. B {\bf 46}, 4667 (1992).

\bibitem{Kubo_coll4}
X. F. Wang and P. Vasilopoulos, 
Phys. Rev. B {\bf 67}, 085313 (2003).


\bibitem{Kubo_Hall}
P. M. Krstajic and P. Vasilopoulos, 
Phys. Rev. B {\bf 83}, 075427 (2011).


\bibitem{broad1}
M. E. Raikh and T. V. Shahbazyan, 
Phys. Rev. B {\bf 47}, 1522 (1993).


\bibitem{streda}
P. Streda,
J. Phys. C {\bf 15}, L717 (1982).

\bibitem{self1}
T. Ando, A. B. Fowler, and F. Stern, Rev. Mod. Phys. {\bf 54}, 437 (1982).

\bibitem{self2}
C. Zhang and R. R. Gerhardts, Phys. Rev. B {\bf 41}, 12850 (1990).





\end{thebibliography}
\end{document}